
\input harvmac

\Title{\vbox{\baselineskip12pt\hbox{Univ.Roma I, 1078/94}
		\hbox{cond-mat/9412069}}}
{\vbox{\centerline{Weighted Mean Field Theory}
	\vskip2pt\centerline{for the}
	\vskip2pt\centerline{Random Field Ising Model}}}

\centerline{David
Lancaster\footnote{$^{(a)}$}{\tt(djl@liocorno.roma1.infn.it)},
Enzo Marinari\footnote{$^{(b)}$}{\tt(marinari@ca.infn.it)},
Giorgio Parisi\footnote{$^{(c)}$}{\tt(parisi@roma1.infn.it)}}
\bigskip
\centerline{(a,c): Dipartimento di Fisica and INFN,}
\centerline{Universit\`a di Roma I {\it La Sapienza},}
\centerline{Piazza A.~Moro 2, 00185 Roma.}
\medskip
\centerline{(b): Dipartimento di Fisica and INFN,}
\centerline{Universit\`a di Cagliari,}
\centerline{Via Ospedale 72, 09100 Cagliari.}

\vskip .3in
Abstract: We consider the mean field theory of the Random
Field Ising Model obtained
by weighing the many solutions of the mean field equations
with Boltzmann-like factors. These solutions are found numerically
in three dimensions and
we observe critical behavior arising from the weighted sum.
The resulting exponents are calculated.

\Date{12/94}

\lref\reviews{Nattermann T and Villain J, Phase Transitions {\bf 11} (1988)
5\semi
Belanger  D P and Young A P, J.~Magn.~Magn. {\bf 100} (1991) 272.}
\lref\GMP{Guagnelli M, Marinari E and Parisi G, J. Phys. {\bf A26} (1993)
5675.}
\lref\MezYou{M\'ezard M and Young P, Europhys.~Lett. {\bf 18} (1992) 653.}
\lref\Grod{M\'ezard M, Talk at the 16th Gwatt workshop on Magnetism, (1992)}
\lref\Remi{M\'ezard M and Monasson R, Phys.~Rev. {\bf B50} (1994) 7199.}
\lref\Aharony{Aharony A, Phys.~Rev. {\bf B18} (1978) 3318.}
\lref\Ogielski{Ogielski A T, Phys.~Rev.~Lett. {\bf 57} (1986) 1251.}
\lref\Sourlas{Sourlas N,  Talk at ``Disordered Systems and Information
Processing'', Heraklion (1994)}
\lref\ImryMa{Imry Y and Ma S K, Phys.~Rev.~Lett. {\bf 37} (1975) 1399.}
\lref\Imbrie{Imbrie J Z, Phys.~Rev.~Lett. {\bf 53} (1984) 1747\semi
 Imbrie J Z, Commun.~Math.~Phys. {\bf 98} (1985) 145.}
\lref\Young{Aharony A, Imry Y and Ma S K, Phys.~Rev.~Lett. {\bf 37} (1976)
1364\semi
 Young A P, J.~Phys. {\bf C10} (1977) L275.}
\lref\PandS{Parisi G and Sourlas N, Phys.~Rev.~Lett. {\bf 43} (1979) 744.}
\lref\Grest{Yoshizawa H and Belanger D P, Phys.~Rev. {\bf B30} (1984)
5220\semi
 Ro C, Grest G S, Souloulis C M and Levin K, Phys.~Rev. {\bf B31} (1985) 1682.}
\lref\Mez{Mezard M, Private communication.}
\lref\Rieger{Rieger H and Young A P, J.~Phys. {\bf A26} (1993) 5279.}
\lref\OglandHuse{Ogielski A T and Huse D A, Phys.~Rev.~Lett. {\bf 56} (1986)
1298.}
\lref\YandN{Young A P and Nauenberg M, Phys.~Rev.~Lett. {\bf 54} (1985) 2429.}

\newsec{Introduction}

Despite progress since the early 1980's, when it was realised that
perturbative techniques fail to capture its critical behavior,
the Random Field Ising Model (RFIM)\reviews\
is still in need of further illumination.
The cause of the difficulty in dealing with this type of disorder is that
the energy landscape
is complicated, with many local minima that pertubation theory fails
to take account of.
Mean field theory provides a simple insight into the difficulty:
at sufficiently low temperature
the mean field equations have many solutions.
The intuitively obvious way of defining mean field theory
would be to weigh these solutions  according
to their Boltzmann factors.
This weighted mean field theory, or
to be precise, the theory with the weights $e^{-\beta F}$ which
include an entropy factor, is the subject of this paper.
In this context the
failure of the pertubative approach comes about because,
as is clear in the supersymmetric formulation\PandS,
the prescription for the weights does not contain
a Boltzmann factor.

In this report we solve the mean field equations numerically
and directly construct the weighted mean field theory.
Our main interest is to see how critical
behavior can arise at the ferromagnetic transition.
Guagnelli {\it et al.}\GMP\ have analysed a truncation of
the theory by only considering the maximal
solutions: those with maximum or minimum magnetisation.
They found that critical behavior does not arise in that case.
Here we shall enlarge on their work by considering a much larger number of
solutions and shall see how divergences in thermodynamic quantities
can arise.
The number of solutions included in the sum is clearly important in this
approach  and we shall discuss the issue in detail.

Replica theory is the natural setting for dealing with the difficulty
mentioned above and although  not yet
successful in treating the critical behavior,
it is helpful to keep the resulting scenario\Grod\ in mind.
The replica symmetric solution becomes  unstable as one reduces the
temperature below $T_{RSB}$, at which point the
correlation length remains finite\Remi. The critical temperature $T_C$ at
which ferromagnetic order arises is at a lower temperature $T_C < T_{RSB}$.
Finally one expects replica symmetry to be restored at an even lower
temperature. This scenario has clear analogues in our mean field treatment.

We start in sections 2 and 3 by discussing the consequences of weighted mean
field theory and describing the technique of numerically searching for
solutions. Section 4 is concerned with the general properties
 and number of solutions we find.
Analysis of the correlation functions and the emergence of
critical behavior is the subject of
section 5. Finally we present a short conclusion discussing
the numerical values of the exponents we determine.

\newsec{The RFIM and its Weighted Mean Field Theory}

The system consists of Ising spins subject to quenched random fields $h_i$;
\eqn\rifm{H = - \sum_{\langle i,j>} S_i S_j - \sum h_i S_i.}
The effect of the fields is to destroy the tendency to long range order,
and the original argument of Imry and Ma\ImryMa\ based
on the energy balance for domain formation gives the lower critical
dimension as  $d_{lcd} = 2$.
Perturbative methods, which lead to the phenomenon of dimensional
reduction\refs{\PandS,\Young}, would instead predict $d_{lcd} =3$.
However in three dimensions it has been shown
rigorously that long range order prevails at low enough temperature\Imbrie.
This case is most interesting and we shall restrict ourselves to
studying the system in three dimensions.

In the work described here the fields are taken to be $\pm |h|$ with
equal probability. According to arguments of Aharony\Aharony, and consistent
with the zero temperature simulations of Ogielski\Ogielski, a bimodal
probability distribution of this sort causes the phase transition to be
first order when $|h|$ is very large.  As in \GMP\ we work with $|h| = 1.5$,
which
is small enough to give a continuous transition, yet large enough
to shift $T_C$ substantially from its value without disorder, thus revealing
a range of non-trivial critical behavior before
the crossover to pure behavior at higher temperature.

For a given realisation of the random fields,
the mean field equations for the system are:
\eqn\mfeqn{m_i = {\rm tanh} \left(\beta (D m_i + h_i) \right), }
where $m_i$ is the local magnetisation and $Dm_i$ indicates a sum of
the magnetisations over the nearest neighbours to site $i$.
As has already been indicated, at low temperature there can be many solutions
to this equation. We shall denote the solutions by $m_i^\alpha$, and
in general shall use the superscript $\alpha$ to denote a quantity,
such as the free energy $F^\alpha$, calculated for that solution.
\eqn\FreeE{ F^\alpha =  V \left(E^\alpha - {1\over \beta}S^\alpha \right), }
where the energy $E^\alpha$, and entropy $S^\alpha$, of the solution are given
by the expressions,
\eqn\EneandEntr{\eqalign{
E^\alpha &=
- {1\over V} \sum_i {1\over 2}m_i^\alpha D m_i^\alpha + h_i m_i^\alpha     \cr
S^\alpha &=
- {1\over V} \sum_i
{1+m_i^\alpha\over 2} \log\left({1+m_i^\alpha\over 2}\right)
+ {1-m_i^\alpha\over 2} \log\left({1-m_i^\alpha\over 2}\right).  \cr
}}

A convenient way of organising the weighted mean field theory
is in terms of a mean field
partition function defined as:
\eqn\mfZ{ Z_{MF} = \sum_\alpha e^{-\beta F^\alpha}. }
This way of writing the theory
simply encodes the intuitive definitions of
quantities, such as the energy, as Boltzmann weighted sums over the solutions.
\eqn\energy{ {1\over V} \langle E \rangle  =
- {1\over V} {\partial \log Z_{MF} \over \partial \beta} =
- {1\over V} \sum_\alpha {\partial  \over \partial \beta}
\left(\beta F^\alpha \right)
 { e^{-\beta F^\alpha} \over Z_{MF}} =
\sum_\alpha E^\alpha { e^{-\beta F^\alpha} \over Z_{MF}} =
\sum_\alpha w^\alpha E^\alpha ,
 }
where we have defined the weights: $w^\alpha = e^{-\beta F^\alpha} /Z_{MF}$.
Note that  $\partial F^\alpha / \partial \beta$ is simple only because
of the constraints
$\partial F^\alpha / \partial m_i^\alpha = 0$,
which are nothing other than the mean field equations \mfeqn.

The average over the random fields is performed as the last step and is
denoted by an overbar, for example $\overline{\langle E \rangle}$.

Critical behavior becomes apparent through study of the
correlation functions. In the RFIM the correlators
$\langle S_i S_j\rangle $ and $\langle S_i\rangle  \langle S_j\rangle $ are
both more singular
in momentum space than the
connected correlator $\langle S_i S_j\rangle _C = \langle S_i S_j\rangle  -
\langle S_i\rangle  \langle S_j\rangle $.
Although different from the field theory usage, it seems common
in random field systems to call the correlator $\langle S_i\rangle  \langle
S_j\rangle $
``disconnected'' and we shall follow this convention. In section 5
we shall determine the exponents $\bar \eta$ and
$\eta$ associated with the disconnected and connected correlators.

Contrary to ordinary mean field theory, the probability distribution
implied by $Z_{MF}$ is not factorised.
The fluctuation dissipation theorem (FDT) is therefore
not needed to calculate the correlation functions.
Using $Z_{MF}$ we find:
\eqn\dissC{ \langle S_i\rangle  \langle S_j\rangle  =
(\sum_\alpha w^\alpha m_i^\alpha ) (\sum_\gamma w^\gamma m_j^\gamma ) }
\eqn\connC{ \langle S_i S_j\rangle _C =
(\sum_\alpha w^\alpha m_i^\alpha  m_j^\alpha )
- (\sum_\alpha w^\alpha m_i^\alpha ) (\sum_\gamma w^\gamma m_j^\gamma )
+ {1\over \beta}  \sum_\alpha w^\alpha g_{ij}^\alpha , }
where $g_{ij}^\alpha$ is defined as $\partial m_i^\alpha / \partial h_j$
and is the usual term arising from the FDT.

\newsec{Solving the Mean Field Equations}

\subsec{Iteration Technique}

Starting from some seed configuration $m^{(0)}_i$,
the mean field equations \mfeqn\ are solved by iteration\Grest:
\eqn\iteqn{m^{(t+1)}_i = {\rm tanh} \left(\beta (D m^{(t)}_i + h_i)\right). }
This is implemented as an efficient code running on the APE parallel
processor.
We insist on strict convergence requirements, that for each solution
$\sum_{board} (m^{(t+1)}_i - m^{(t)}_i)^2 < 2.10^{-13}$. Where the
$m_i$'s are represented to float accuracy and there are 128  APE boards
to cover the complete $32^3$ lattice. With this requirement we find no
difficulty in distinguishing  solutions.
The criterion for saying that two solutions are the same,
that the maximal site difference $|m_i^\alpha - m_i^\gamma|$,
be less than some cutoff,  leads to the same identification of
solutions for a wide range of cutoff values. We have finally
chosen this cutoff to be $10^{-3}$.
Although the majority of seeds quickly converge, for some
temperatures and some realisations of the random field we
found that the convergence time was unacceptably large.
A maximum number of 15000 iteration steps has been imposed, leading
to rejection of  about 1.4\% of the potential solutions.
We have observed a phenomenon known as ``funneling'' in which
the configurations arising from different seeds
rapidly converge to very similar configurations which then follow
the same path though configuration space before finally converging\Mez.
The iteration technique may lead to solutions that are
maxima of the free energy besides minima. Although we do
not see such solutions in the cases described below,
we have no simple way of rejecting them and
they are implicitly included in the weighted sum.

\subsec{Maximal Solutions}

The iteration \iteqn\ has the property that if one can assign an ordering
to two configurations $m_a$ and $m_b$: $m^{(t)}_{a\ i} \ge m^{(t)}_{b\ i}$
for all sites $i$;
then this ordering is preserved: $m^{(t+1)}_{a\ i} \ge m^{(t+1)}_{b\ i}$.
Consequently one can identify two special, maximal solutions,
$m_+$ and $m_-$, that arise
from the seeds, $m^{(0)}_{+i} = 1$ and $m^{(0)}_{-i} = -1$, and that bound any
other solution since
$m^{(t)}_{+i} \ge m^{(t)}_{any\ seed\ i} \ge m^{(t)}_{-i}$.
These were the solutions analysed in \GMP.

The existence of maximal solutions provides an accurate means of searching for
the temperature (which by analogy with the replica analysis,
we denote $T_{RSB}$) at which the mean field equations start to
have more than one solution. Below this temperature $m_+$ and $m_-$ differ.
We have found $T_{RSB}$ for series of random fields
both on $32^3$ and $64^3$ lattices. At the larger size the peak
of the distribution of $T_{RSB}$'s moves to higher temperature
while the width shrinks -- supporting the results from replica theory
that $T_{RSB}$ is well defined and separated from $T_C$
in the thermodynamic limit.
In about 10 percent of the cases we examined
we observed two or more values of $T_{RSB}$. That is:
at high temperature there is a unique solution, then as
the temperature is reduced we first find
more than one solution, then an interval where again only one
solution exists, before finally reaching another low temperature region
with many solutions. This effect is reminiscent of some of the observations
of jumping made by Sourlas at zero temperature\Sourlas.

\subsec{Seed Strategies}

We have considered several strategies
for the choice of seeds.
The most efficient strategy we found was to
consider chequerboard seeds: we divide the $32^3$ cube
up into 8 subcubes of size $16^3$, and colour each one independently with $+1$
or $-1$. The leads to 256 seeds including the maximal ones.
Further subdivision was thought to be impractical, so further
sets of 256 seeds were generated by adding independent random pertubations
to each site of the chequer seeds.

\newsec{Number and Properties of Solutions}

The number of solutions found and included in the weighted mean
field theory is an important parameter. We start this section
by displaying the average number and the effective number
of solutions we find in our work.
The main justification however, for curtailing the search for
further solutions at the point we choose, is that the quantities
measured from the correlators show very little
change on increasing the number of solutions beyond those found with
chequerboard seeds alone.

The results we shall present are based on data for the $32^3$ periodic lattice
for a set of 150 different magnetic field samples.
We have calculated quantities based on
the maximal solutions alone, and the solutions obtained from chequerboard
seeds.
Besides these, in the temperature region identified as most interesting,
we have performed
longer runs including chequerboard followed by two sets of
chequerboard plus random seeds
for the first 100 of the magnetic field samples.
The temperature ranges covered by each of these three data sets
are most clearly seen by looking ahead to the ranges of the three curves of
fig. 2.

In all cases the temperature is
measured in units of the mean field critical temperature for the pure model
$T/T_{C\ pure} = T/6$.

Quantities such as the Energy, Entropy and Free Energy vary
smoothly with temperature and the curves show little variation
when the number of solutions included is increased.
The average squared site magnetisation is also a smooth function, but
the average magnetisation itself suffers large fluctuations.
The specific heat starts to develop a cusp at the temperatures we
shall later identify as the ferromagnetic phase transition.

\subsec{Number of solutions}

The number of distinct solutions found by iteration is shown in
\fig\Nsolns{Total number of solutions found. Squares for chequerboard
seeds alone, triangles for chequerboard followed by one set of random
pertubations added to chequerboard, open triangles for
chequerboard followed by two such sets of random pertubations.}.
Those solutions with completely
negligible weight ($F^\alpha - F_{min} > 80$) have been dropped.
The three sets of points are from successively increasing numbers of
iteration trials. The
lowest set (squares) is for the 256 chequerboard seeds while the upper
 two sets (triangles)
correspond respectively to the
addition of one and two sets of
random pertubations on top of chequerboard.

Already from this figure it is clear that there is a law of diminishing
returns relating the number of solutions found to the number of seeds
iterated. From more detailed studies at a fixed low temperature, we see
that after a steep initial rise, the number of solutions only increases slowly
with the amount of effort. We have determined to stop searching for
solutions at the point determined by the shoulder of this curve, which
in practice is after the chequerboard and two sets of chequerboard
plus random seeds.
At higher temperatures a less though search is sufficient.

A feeling for the degree to which extra solutions are important
comes from taking account of their weights, and a convenient quantity
is $W^2$ defined by:
\eqn\Wsq{W^2 = \sum_\alpha (w^\alpha)^2 .}
The inverse gives the effective number of solutions that
contribute. A plot of $W^2$ is shown in
\fig\Wsqfig{$W^2 = \sum_\alpha (w^\alpha)^2$, the inverse of the
effective number of solutions. Circles are for maximal solutions alone,
squares for chequerboard, and triangles for chequerboard followed by
(2 sets of) additional pertubations.}
for maximal, chequerboard and chequerboard plus random seeds.
At high temperatures $W^2 = 1$ since only one solution exists,
whereas the rise at low temperatures is due to the increasing
dominance of certain solutions.
This increasing dominance can also be seen  in the density
of solutions in free energy which rises less steeply
at lower temperatures.
The fact that the effective number of solutions grows more slowly than
the total number of solutions gives us our confidence that we
are finding all the important solutions.

A histogram of the overlaps between different solutions displays a
peak at small overlap with a long tail. When the histogram is
normalised by the product of the weights of the solutions the
peak moves to smaller overlap.

\subsec{Form of solutions}

There is a regime of temperature where it is possible to get some
intuition into the form of the solutions.  Generally the solutions are
complicated: at high temperature $m_i$ is small but follows the
field $m_i \approx \beta  h_i + \beta^2 D h_i$, whereas at low
temperatures there is an overall magnetisation and sometimes
reversed field domains. However, in the region just below $T_{RSB}$ where there
are only a small number of solutions, it is possible to see that the
differences between solutions are local. In \fig\Blobs{Local
differences between maximal solutions at temperatures just below $T_{RSB}$,
isosurfaces of difference $m_{+i}-m_{-i} = 0.1$.}
a three dimensional picture of
the difference in magnetisation between the maximal solutions (note that this
is always positive) is shown
by the constant difference surfaces. The dark region has small
difference. The notable feature is that the regions of appreciable
difference do not touch each other and are separated by a sea where the
solutions are almost identical.
This suggests that when the difference between
maximal solutions has $N$ local regions, each can be independently
switched to either of two configurations, and that there should be a total
of $2^N$ solutions. In fact, in the example corresponding to \Blobs, we do find
a total of 8 solutions, and this method has been used to test the
efficiency of different seed strategies.
These solutions are found
where the regions are local and do not influence each other.
At such temperatures we must expect a constant
density of such local differences between maximal solutions, and thus
a total number of solutions that grows exponentially with the volume of the
system.
This picture fails at lower temperatures where the difference between maximal
solutions is no longer local.

As was the case for the maximal solutions\GMP, in any
particular solution correlation functions are rather rough
with finite correlation length.

\subsec{Below the transition}

The reader will have noticed that our definition of mean field theory
as a sum over {\it all} states is at variance with our usual understanding of
the the ferromagnetic phase transition. One might try to modify the
definition at low temperature by only including solutions with one
sign of the magnetisation, however the region in the vicinity of the
transition will remain unclear.
It is possible to visualise how the transition takes place
from the lists of solutions along with their magnetisations and weights.
 In the top
row of \fig\Transition{Number (top line) and weight (bottom line) of
solutions shown as a histogram against magnetisation for temperatures
decreasing towards the right.}
the number of solutions is shown as a histogram against their
magnetisation. The width of the distribution increases at lower temperature,
but there are always solutions of small magnetisation. In the lower
line of plots in \Transition\ the summed weights  rather than the number
of solutions are shown. It is clear that at low temperature the
small magnetisation solutions have low weight and are unimportant,
and that the space of solutions is divided into two significant groups
with opposite magnetisation.

\newsec{Critical Behavior}

In order to study critical behavior of the theory we look at the
susceptibilities and the
quantities that arise from analysing the correlation functions.
Besides investigating the divergence of the
correlation length, we expect two independent exponents defined
by the behavior at the critical point.
In three dimensional real space the disconnected case goes as
$\sim r^{1 - \bar \eta}$
whereas the connected case dies more quickly as
$\sim r^{-1 - \eta}$.

We always consider plane--plane correlators in one spatial direction
with the transverse momenta set to zero:
$G(x,k_y=0,k_z=0)$.
{}From the solutions $m^\alpha_i$ two different
lattice correlation functions, $C^{(1)}$ and $C^{(2)}$,
can be defined using sums over transverse
planes. For example in the $x$ direction:
\eqn\Planesum{ P^\alpha_x(x) = {1\over L_y L_z}\sum_{y,z} m^\alpha(x,y,z). }
For $C^{(1)}$, which is the plane--plane version of the
disconnected correlator \dissC,
the weighted sum over solutions is,
\eqn\dissClat{\eqalign{
C^{(1)\alpha}_x(x) &=
{1\over L_x}\sum_{x_0} P^\alpha_x(x_0) P^\alpha_x(x_0 + x)  \cr
C^{(1)}_x(x)         &=
\sum_{\alpha} w^\alpha C^{(1)\alpha}_x(x).  \cr}}
Instead of performing the sums
in the order given above, we can
first weight the plane sums before combining them,
\eqn\otherClat{\eqalign{
P_x(x)             &=
\sum_{\alpha} w^\alpha P^\alpha_x(x) \cr
C^{(2)}_x(x)       &=
{1\over L_x}\sum_{x_0} P_x(x_0) P_x(x_0 + x). \cr } }
This procedure yields the plane--plane version $C^{(2)}$ of the
first term in the connected correlator $\langle S_i S_j\rangle _C$, \connC.
An average over the principal directions and over the random fields
is made before obtaining the final results
for  $\overline{C}^{(1)}$ and  $\overline{C}^{(2)}$.
The plane--plane connected correlator is given by
$C^{(2)}-C^{(1)} + C^{(FDT)}$,
in which the FDT term cannot be determined
directly from the solutions $m^\alpha_i$ and a further iteration is
required.

\subsec{Disconnected Correlator, $\langle S_i\rangle  \langle S_j\rangle $}

Scaling arguments suggest that away from the critical point
the 3 dimensional disconnected
correlator behaves as $G({\bf r}) = r^{1-\bar\eta} f(r/\xi)$.
The scaling function we shall use to fit $C^{(1)}$
and define the correlation length arises
from a Lorentzian squared propagator,
a form that is motivated by zeroth order pertubation theory.
This choice is recommended by its simplicity and it
is frequently employed for fitting experimental data.
The disadvantage of this choice of scaling function
$f(r/\xi) = (\xi/r)^{1-\bar\eta} e^{-r/\xi}$,
is that the critical limit, $\xi \to \infty$,
amounts to $\bar\eta = 1$ behavior.
Although  $\bar\eta$ is in fact close to 1, we have explicitly
checked that the other natural choice of scaling function
$f(r/\xi) = e^{-r/\xi}$, gives similar results.

The plane--plane form of our fitting function is:
\eqn\fitting{  A \left(1 + {x \over\xi}\right) e^{-x/\xi} + B.}

In contrast to the lack of sensitivity to the precise form
of scaling function, it is important to make the periodic
modification correctly.
Since the correlations fall off so slowly, only
keeping the leading two terms in the periodic sum leads to significant
errors. We use the complete form:
\eqn\fittingper{\eqalign{
  &{A \over (1-e^{-L/\xi}) }  \left(
\left(1 + {x \over\xi}\right) e^{-x/\xi} +
\left(1 + {L-x \over\xi}\right) e^{-(L-x)/\xi} \right) \cr
& \quad + {A e^{-L/\xi} \over (1-e^{-L/\xi})^2 } {L\over \xi} \left(
 e^{-x/\xi} + e^{-(L-x)/\xi} \right)
\quad+\quad B . \cr} }

In all cases, even well away from the critical point,
the fits are extremely good.
Although we work exclusively above the critical temperature
a constant term $B$ has been included.
This is done
in order to take account of finite size effects
and is also necessary to avoid
discontinuities between data sets from different
number of seeds.
In the infinite system we would expect $B$ to be
the mean site magnetisation squared
$\overline{\rm m}^2$;
we however always observe it to be larger, approaching
$\overline{\rm m^2}$ at low temperature.
We shall regard $B$ becoming different from zero as
a signal of finite size effects being important.

In \fig\MDiss{Constant part of the disconnected correlator $B$,
the squared magnetisation.
Circles indicate results for maximal solutions alone, squares
are for chequerboard solutions.}
and \fig\XiDiss{Correlation Length $\xi$, from the disconnected correlator.
Circles indicate results for maximal solutions alone, squares
are for chequerboard solutions.
The line shows the best fit, ignoring the lowest two points
which are not consistent with $B = 0$.}
the constant $B$ and
the correlation length obtained from the
fitting are plotted
against temperature. The jacknife technique is used for the errors, but note
that they are correlated since the same set of random fields is used
at each temperature.
The plots are displayed for the maximal and for
the chequerboard solutions.
These sets of data flow smoothly into each other
as do the points based on the solutions arising from
chequerboard followed by chequerboard
plus random seeds. In fact these later points lie well within the errors
of the chequerboard points and are not shown.  It is generally
 the case
that extra solutions
beyond chequerboard do not change any of our analysis of
 the disconnected correlator.
For each of the fitting parameters the
 quantity $\overline{x^4} / 3 (\overline{x^2})^2$,
remains close to  1
(its value for a gaussian distribution),
except
for the measurement of the constant term $B$ at high temperatures
where it becomes as large as 2.

Consider \MDiss\ for the magnetisation squared as measured by
the constant term $B$ of the correlator.
Significant deviations between the points calculated
with many solutions and those calculated with only maximal solutions
become apparent below the temperature at which the maximal solution
$B$ becomes non-zero ($T \sim 0.77$).
The value of $B$ calculated
with many solutions remains zero to a lower temperature ($T \sim 0.74$)
before it
too becomes non zero, signalling either the onset of finite size effects
or a breakdown of the approximation.
In this temperature interval where deviations are clear yet the
theory is reliable,
the correlation length calculated from many solutions continues to grow while
that of the maximal solutions rounds off.
Certainly the rounded form coming from the
maximal solutions is not a finite size effect as the curve
for maximal solutions at $64^3$ lies directly over this set.
On the other hand the growing curve
 obtained from many solutions
can become divergent as the size is increased since the number
of solutions included will also increase.
An exponent for this potential divergence can be determined
by fitting the curve $\xi \sim (T-T_C)^{-\nu}$.
To do this, a combination of points from different data sets that include
as many solutions as available, is used.
The two lowest temperature points where $B$ is non-zero are dropped,
while at the other end of the range the fit is found to be insensitive to the
inclusion or not of high temperature points.
We find $\nu$ to be $1.25 \pm 0.11$ where the errors
are statistical from jacknife.
The value of $T_C$ found in this way is $0.64 \pm 0.01$.

The fit to the correlator, specifically the coefficient $A$,
allows an estimate of the
exponent $\bar \eta$. Scaling requires that the correlators
$G(x,k_y=0,k_z=0)$ behave as $\xi^{3-\bar\eta} \tilde G(x/\xi)$,
leading in the case of the fitting function \fitting, to
$A \sim \xi^{3-\bar\eta}$.
To avoid the uncertainty in $T_C$
we plot $\log(A/\xi^3)$ against $\log(\xi)$ in
\fig\ADiss{The parameter $A/\xi^3$, from the disconnected correlator.
Each point corresponds to the data set with most solutions available
at that temperature. The last point is not included in the fit
as $B \neq 0$.}.
In contrast to the situation for the correlation length
this logarithmic plot is curved and the
high temperature points adversely affect the exponent.
The line shown on the figure is obtained as the slope of
the last points which are consistent with $B = 0$.
This provides an estimate
of  $\bar\eta = 0.89$ with a statistical error of $\pm 0.10$.
The systematic error associated with this measurement
is harder to evaluate, but from the tendency of the
lowest temperature points that were omitted, it
is surely in a direction that would increase  $\bar\eta$.

An alternative method of evaluating  $\bar\eta$ would be directly
through the susceptibility:
\eqn\dissX{ \chi^{(1)} = {\beta \over V} \sum_{i,j} \langle S_i\rangle  \langle
S_j\rangle
= {\beta V} \langle m\rangle^2
= {\beta V} \left( {1\over V} \sum_i w^\alpha m^\alpha_i \right)^2 .
}
After the average over disorder we find that the susceptibility
is given in terms of the magnetisation fluctuations
$ {\overline\chi^{(1)}} = {\beta V} {\overline{{\rm m}^2}}$
yielding a curve similar to the one obtained from the
correlator fit.
This curve does not however allow a determination of the
exponent because in the critical low temperature region,
$\overline{{\rm m}^2}$ is contaminated with
parts coming from non-zero magnetisation
causing the slope to actually turn positive
for the points not consistent with $B = 0$.

\subsec{Connected Correlator, }

The analysis of the
connected correlator requires  more delicacy.
There are two reasons for this;
firstly the extra FDT term and
secondly  a strong sample dependence on the random field.
By evaluating the quantity $\overline{x^4} / 3 (\overline{x^2})^2$
for the fitting parameters, one can see
that in contrast to the disconnected case, the distributions are
distinctly non-gaussian. This observation is reinforced by an
inspection of the susceptibilities for individual magnetic fields
where one often observes points far removed from the mean.

The full connected correlator is given by
$C^{(2)}-C^{(1)} + C^{(FDT)}$.
The FDT part $C^{(FDT)}$,
defined in terms of  $g_{ij}^\alpha =\partial m_i^\alpha / \partial h_j$
according to the plane--plane version of \connC,
cannot be determined directly from the magnetisations.
In fact a separate iteration is required to solve the equation
for $g_{ij}^\alpha$:
\eqn\giteqn{g_{ij} = \beta (1 - m_i^2)(D g_{ij} + \delta_{ij}). }
It would be prohibitive to perform this iteration for each
site $j$ of the lattice.
However for a small number of cases we have done the
iteration for one site fixed
and note a strong dependence on the chosen site.
We find that $g_{ij}$  is more singular in
position space than the part $C^{(2)}-C^{(1)}$
 arising from the magnetisation, and
that asymptotically  $g_{ij}$ always goes to zero.
This is consistent with our theoretical prejudice that a
fit of the FDT part alone should be done with the
function arising from a propagator with a single pole in momentum space.
At higher temperatures the FDT piece is dominant in the connected correlator,
at lower temperatures it can plausibly be neglected. In that case,
an analysis of $C^{(2)}-C^{(1)}$ alone leads to
a correlation length,  at the lowest
few temperature points,
that is within the errors of the same quantity,
\XiDiss, calculated for the disconnected correlator.

To determine the exponent $\eta$  it is more convenient and accurate
to abandon the full correlator and to work with the susceptibility.
\eqn\connX{\eqalign{
\chi^{(C)} &= {\beta \over V} \sum_{i,j} \langle S_i S_j\rangle _C =
\chi^{(2)}-\chi^{(1)} + \chi^{(FDT)} \cr
\chi^{(2)}-\chi^{(1)} &=
 {\beta V} \left(
(\sum_\alpha w^\alpha m^\alpha  m^\alpha ) -
(\sum_\alpha w^\alpha m^\alpha )^2 \right) \cr
\chi^{(FDT)}  &=
{1\over V} \sum_{ij}\sum_\alpha w^\alpha g_{ij}^\alpha
= {1\over V} \sum_i\sum_\alpha w^\alpha g_{i}^\alpha. \cr }
}
The advantage of working with the susceptibility is that
$ g_i^\alpha = \sum_{j} g_{ij}^\alpha$
can be evaluated by
a single iteration of the equation obtained by summing  \giteqn:
\eqn\Xiter{g_i = \beta (1 - m^2_i) \left( D g_i + 1 \right). }
The term $\chi^{(FDT)}$ is not
very sensitive to the number of solutions employed to calculate it.
In fact, in the temperature region of interest the FDT term is small compared
with the difference term and is well approximated by
the contribution from the maximal solutions alone.
On the other hand, the magnetisation difference term does show
some dependence on the number of solutions; the dependence is still
small, but leads to different estimates for the exponent.
Both terms, $\chi^{(2)}-\chi^{(1)}$ and $\chi^{(FDT)}$,
and their sum $\chi^{(C)}$ are shown in
\fig\Xconn{Connected Susceptibility, $(\chi^{(2)}-\chi^{(1)})$  part shown
with squares (chequerboard data), FDT piece $\chi^{(FDT)}$ shown with
circles for the maximal solutions only (the errors are small),
the sum $\chi^{(C)}$ shown with open triangles.
The line shows the best fit using only low temperature points.}.
The crossover to pure behavior is very clear
and using only the points below the crossover temperature we
can attempt to determine the exponent $\eta$.
The analog of the method used in the disconnected case, plotting
$\chi^{(C)} /\xi^2$ against correlation length, fails
because the curve turns and begins to
rise at low temperature in the same way as for $\chi^{(1)}$ based on
${\overline{{\rm m}^2}}$
We therefore resort to a direct determination of the susceptibility
exponent $\gamma$ defined by $\chi^{(C)} \sim (T-T_C)^{-\gamma}$.
Because of the small number of points and the large errors
the points are not weighted according to the size of their errors
in this fit.
We obtain $\gamma = 1.66 \pm 0.53$ with chequerboard data.
For the increased number of solutions corresponding to
chequerboard followed by two sets of random pertubations
we find that $\gamma$ is shifted to $\gamma = 1.97 \pm 0.91$,
where the increased size of the error is due the reduced number of
random fields in this sample.
Using the scaling relation $\gamma = \nu(2-\eta)$,
we determine the exponent $\eta$ to be
$\eta = 0.7 \pm 0.5$ and
$\eta = 0.4 \pm 0.8$ for these respective data sets.

\newsec{Conclusion}

By the direct procedure of solving the mean field equations for the RFIM
we have analysed the natural mean field theory obtained by weighting
the solutions with Boltzmann-like factors.
Our main conclusion is that this theory can describe critical behavior
at the ferromagnetic transition. The critical divergences arising
from the sum over many solutions, each of which has regular behavior.
We have investigated both the number and properties of the solutions
that contribute.

The exponents we have determined are,
$\nu = 1.25\pm0.11, \bar\eta = 0.89\pm 0.10,\eta = 0.4 \pm 0.8$
where the errors are statistical.
The most accurate determinations to date are from
the work of Rieger and Young who use
Monte Carlo and finite size scaling \Rieger.
Our estimate of $\nu$ is consistent with their value of
$\nu = 1.4\pm0.2$.
Although our result for $\eta$ lies close to their value of
$\eta = 0.6 \pm 0.1$,
our errors are so large that
this hardly provides a stringent test of our method.
Given the systematic
errors involved in our determination of  $\bar\eta$, which would
tend to increase it, there is no contradiction
with their value of
$\bar\eta = 1.04\pm 0.08$.
As is clear from the decay of the disconnected correlator at criticality,
$\bar\eta$ must be greater than 1 for
consistency of this analysis.
The fact that it is so close to 1
has led to suspicions of a first order transition \Rieger,\YandN.
However we find it impossible to fit the specific heat,
as measured by the derivative of the energy, to a divergence,
and suspect a cusp form instead.
There is thus the puzzle of zero latent heat already noted in \Rieger.
In order to see how this effect in particular, and the analysis
in general, depends on the choice of probability distribution for the
random fields, we have been studying the gaussian distribution.
Preliminary results suggest that conclusions are unaffected.

Only including those solutions that arise from chequerboard seeds
gave good results for the disconnected correlator. Further accuracy
would come from an increase in the number of random field samples rather
than an increase in the number of solutions.
In the connected case we observed stronger
corrections coming from including solutions beyond chequerboard,
however even there it was mainly the sample dependence that prevented
greater accuracy.

\bigbreak\bigskip\bigskip\centerline{{\bf Acknowledgments}}\nobreak
D.~L. would like to thank Remi Monasson for many helpful discussions.

\listrefs
\listfigs
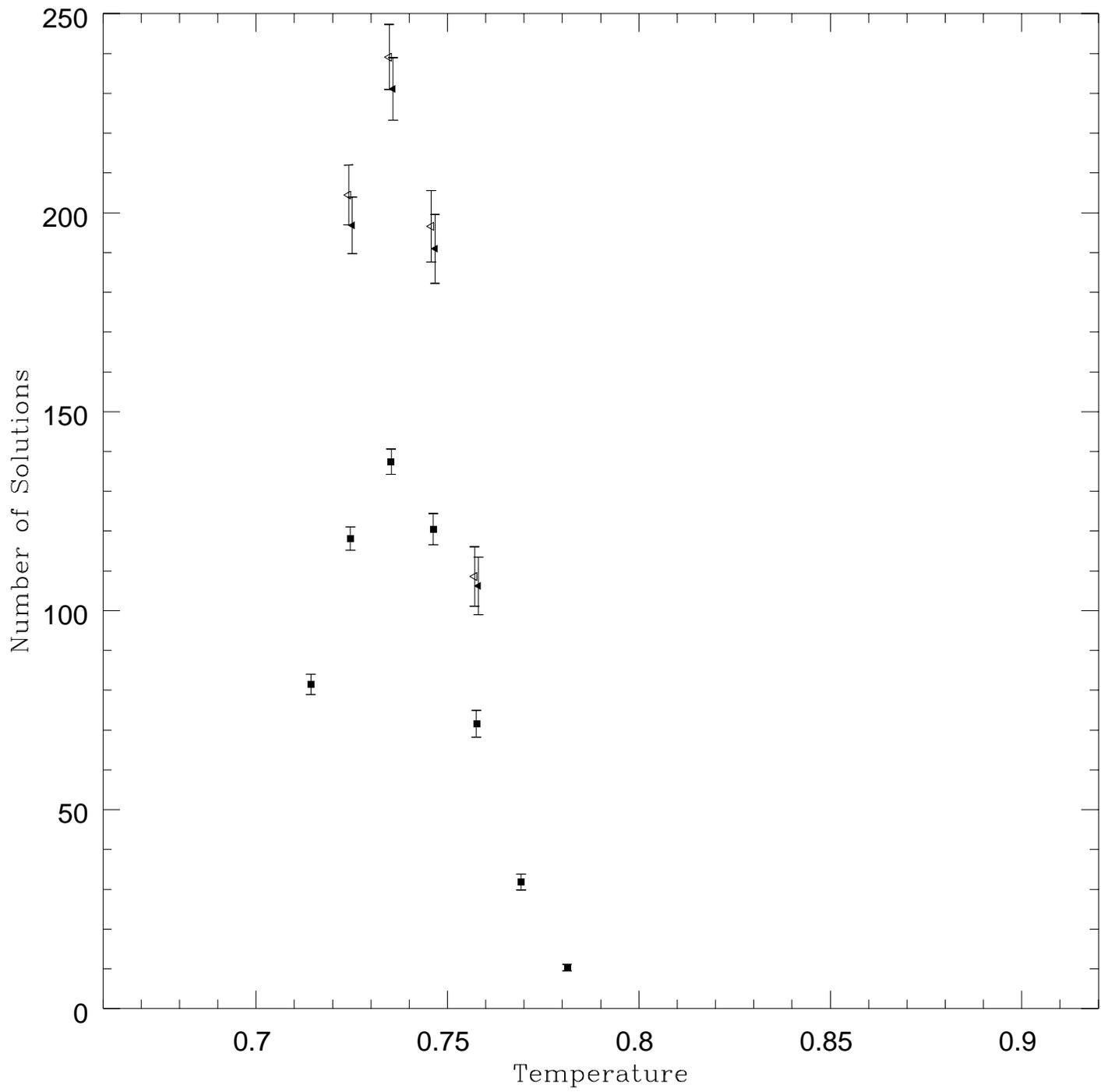
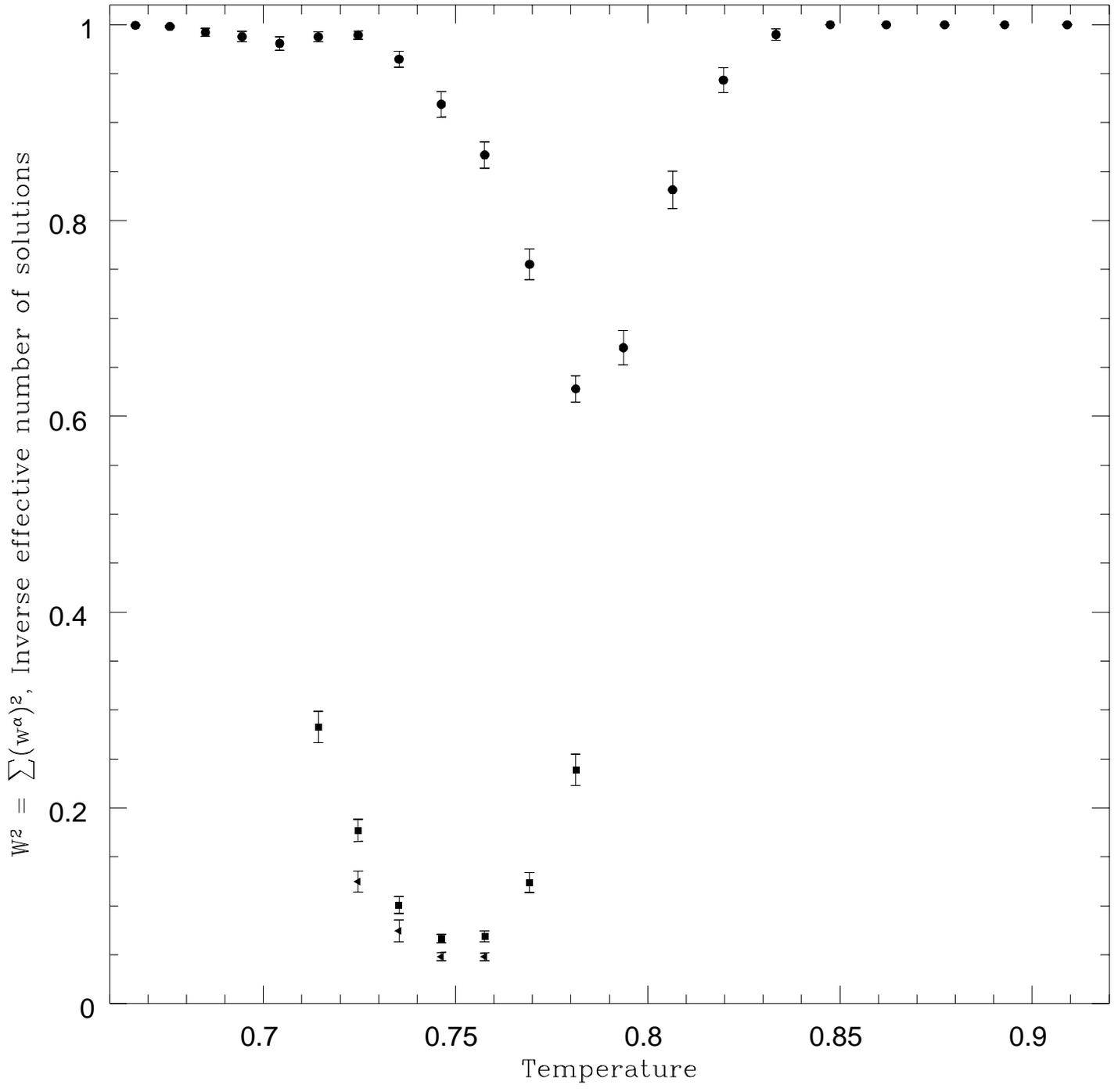
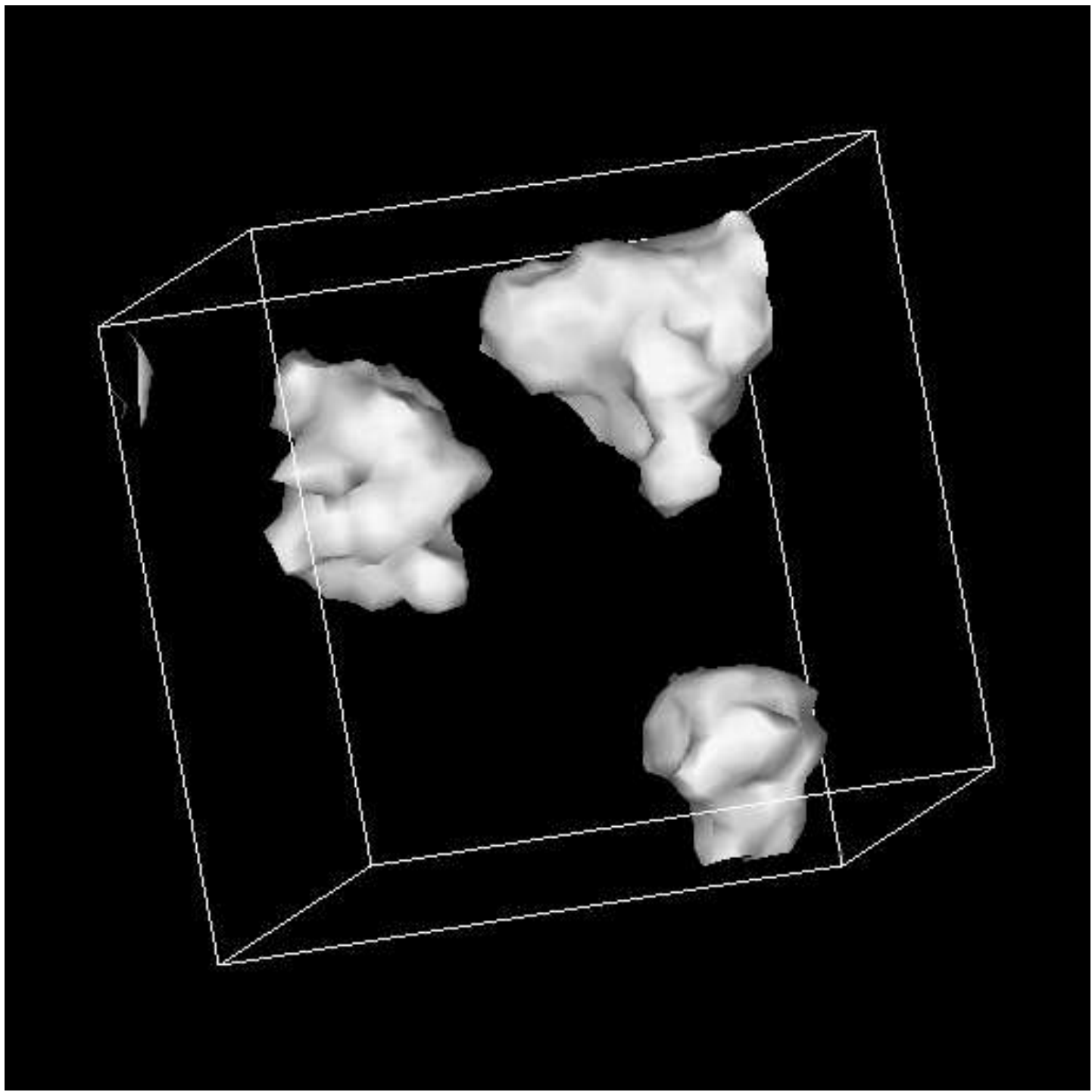
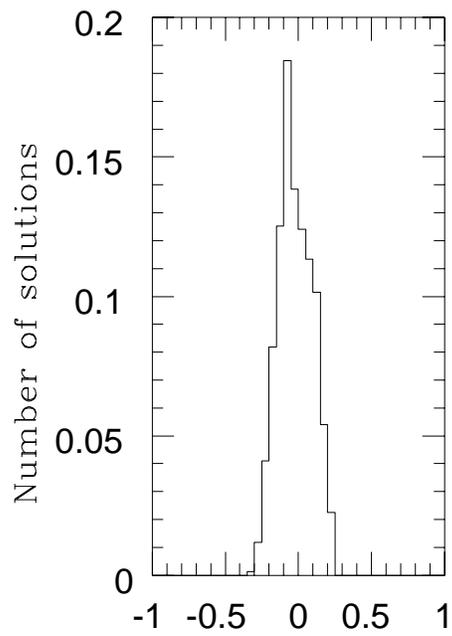
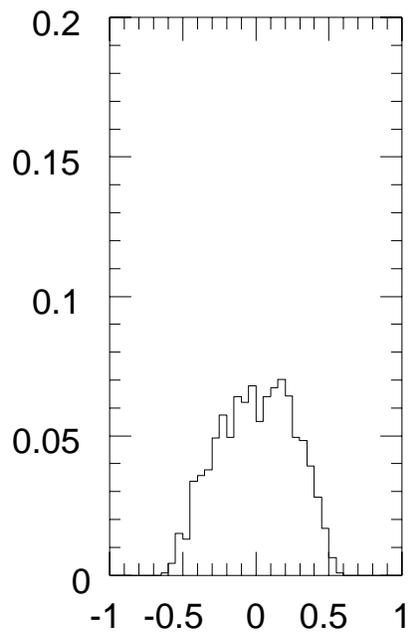
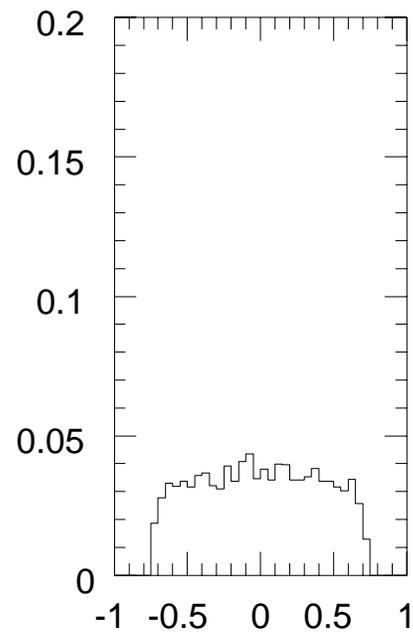
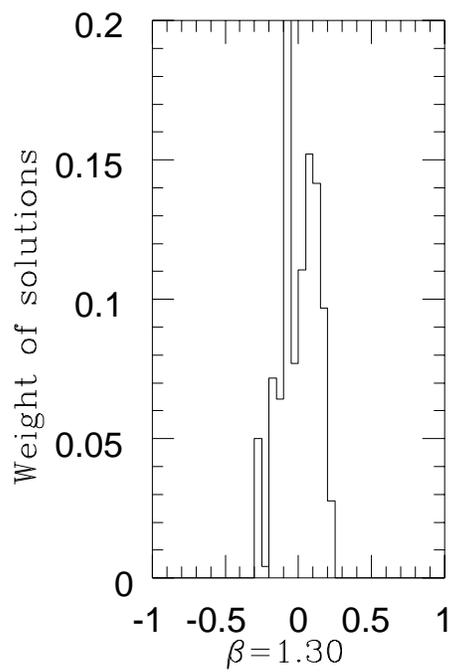
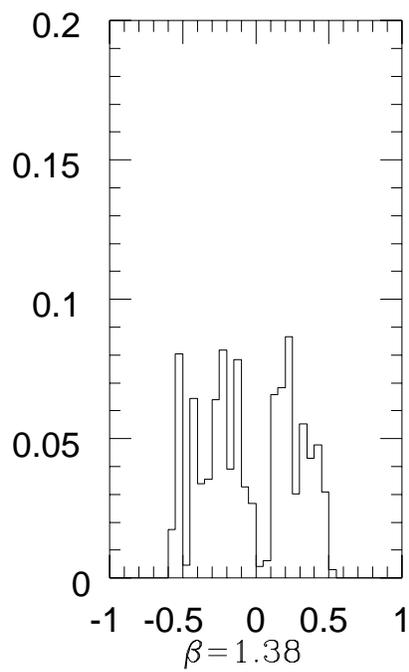
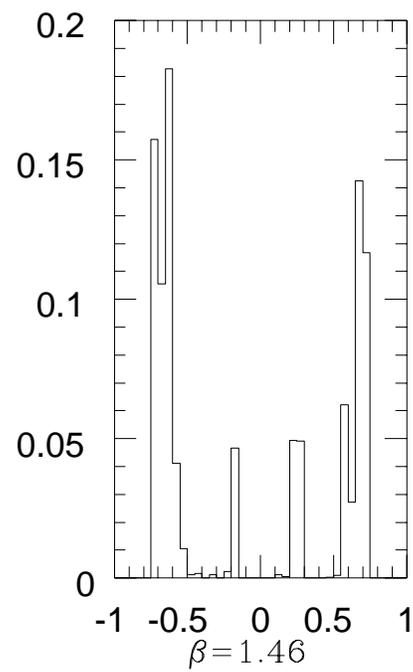
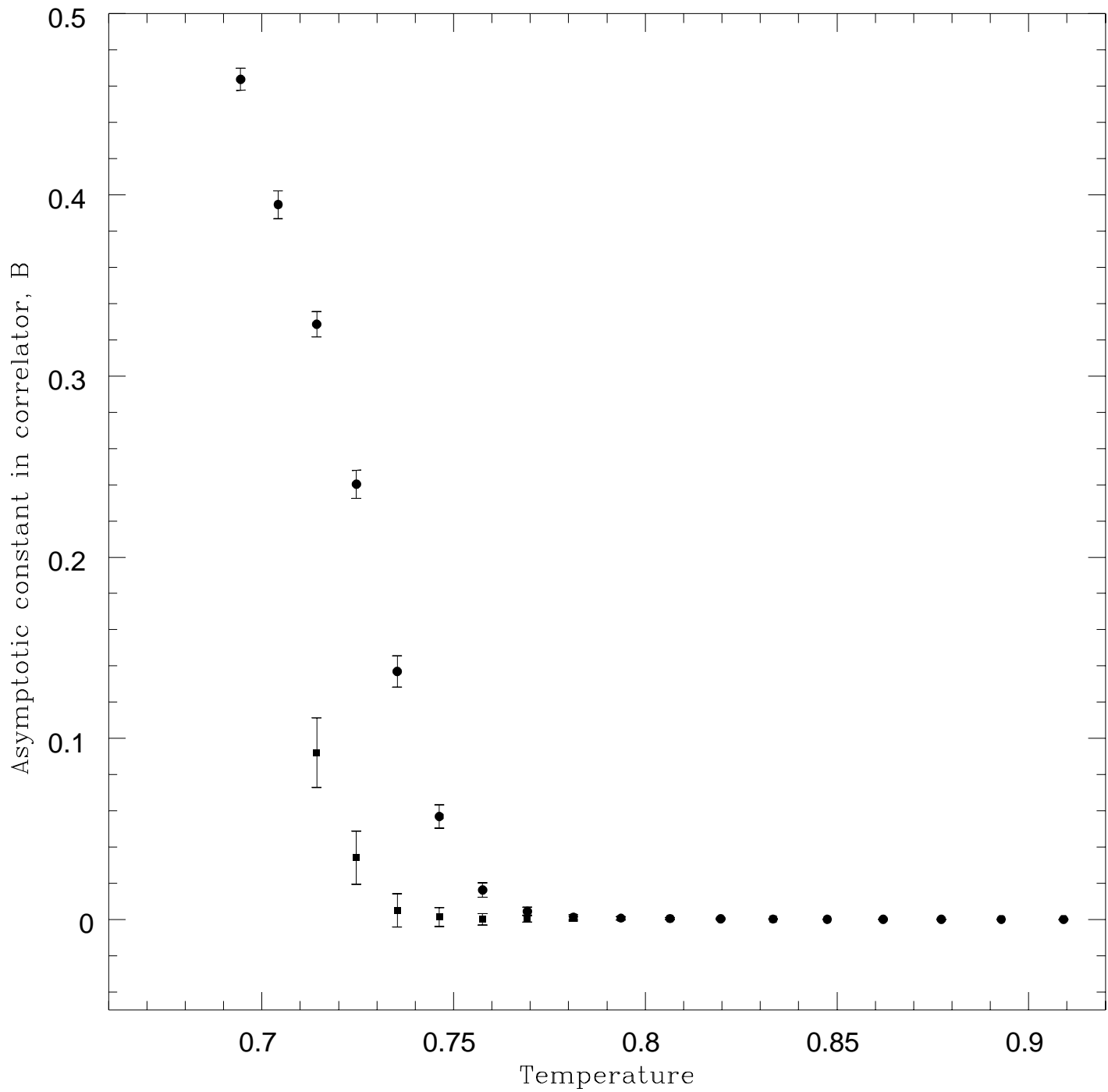
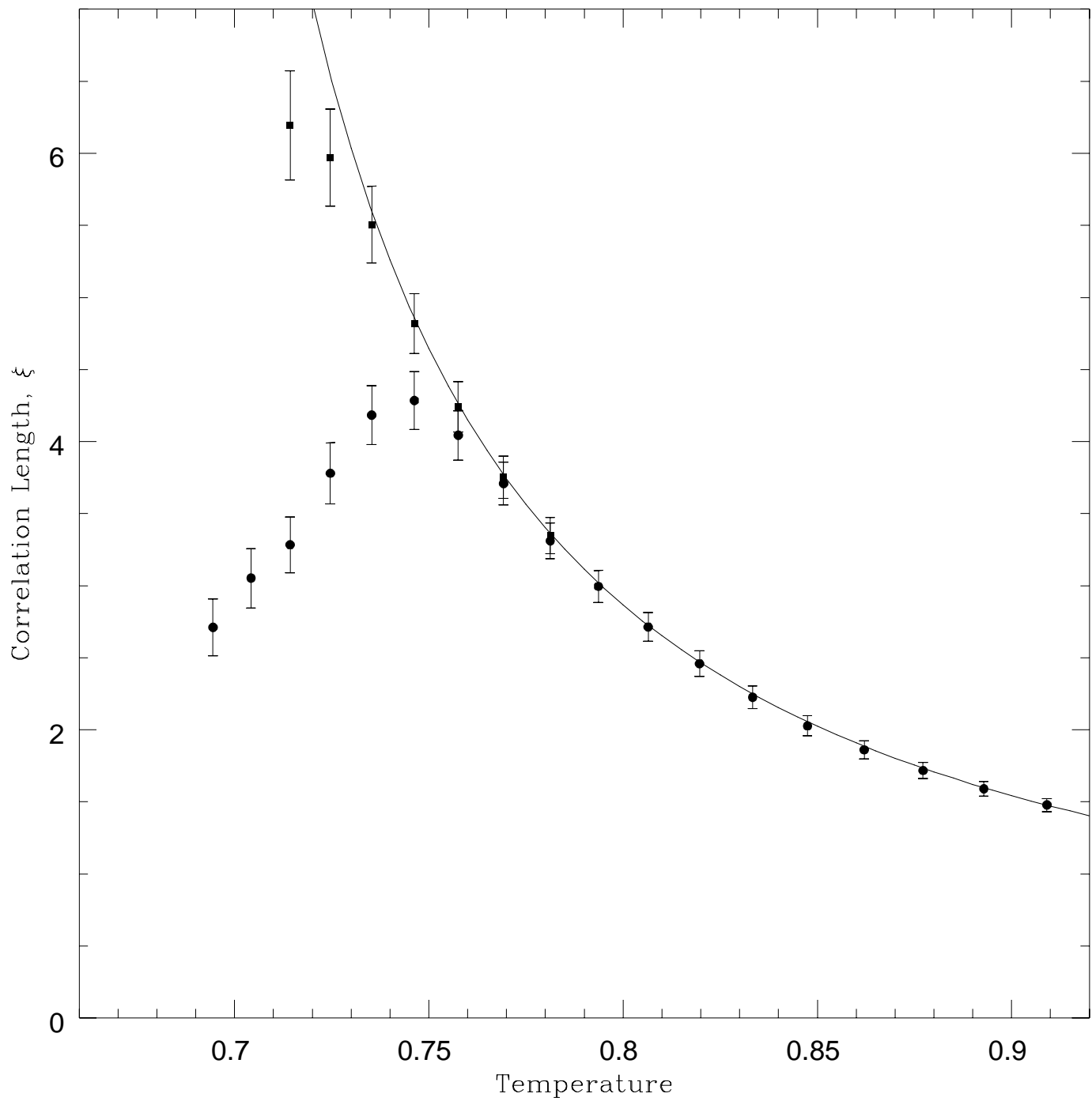
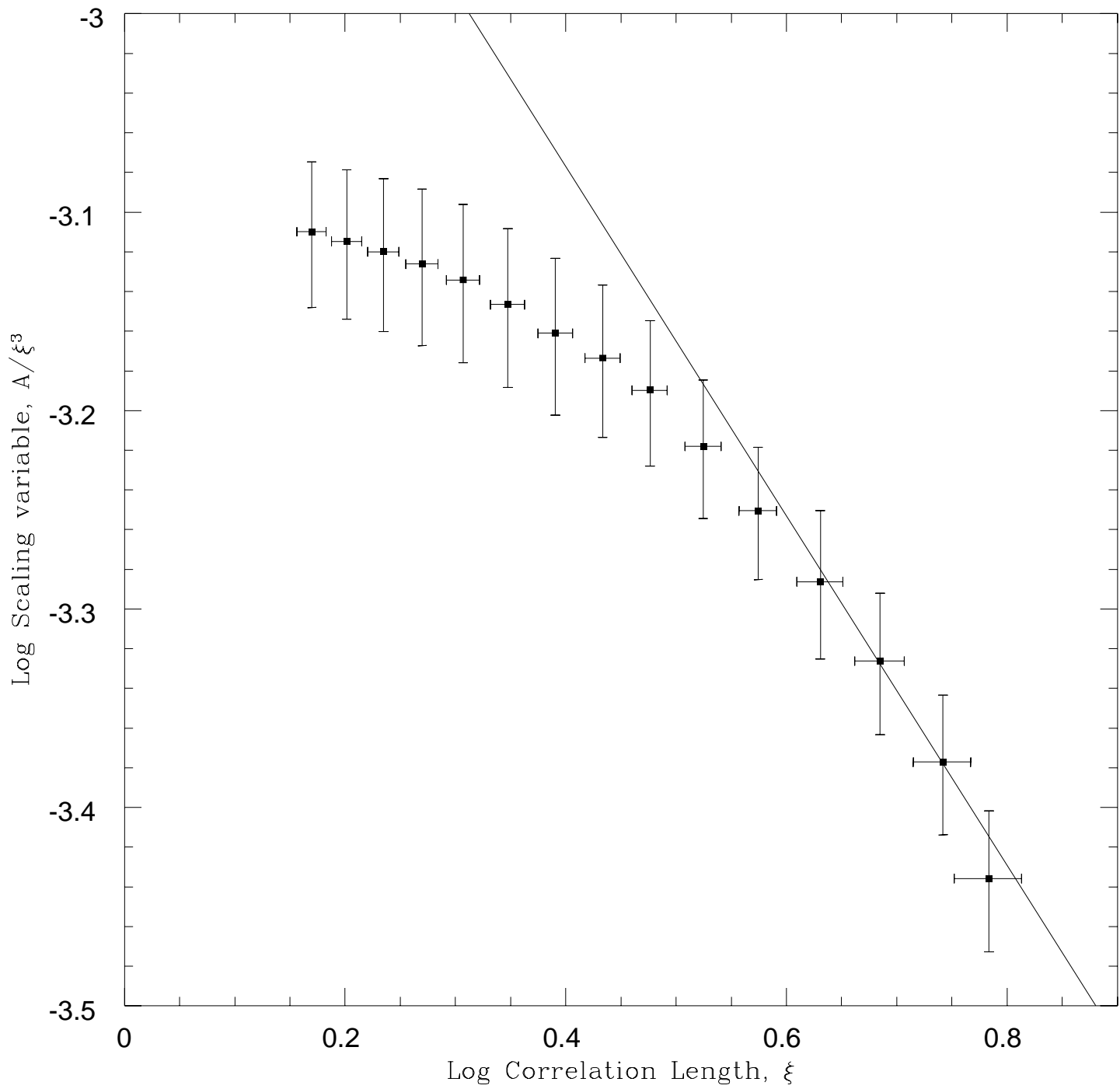
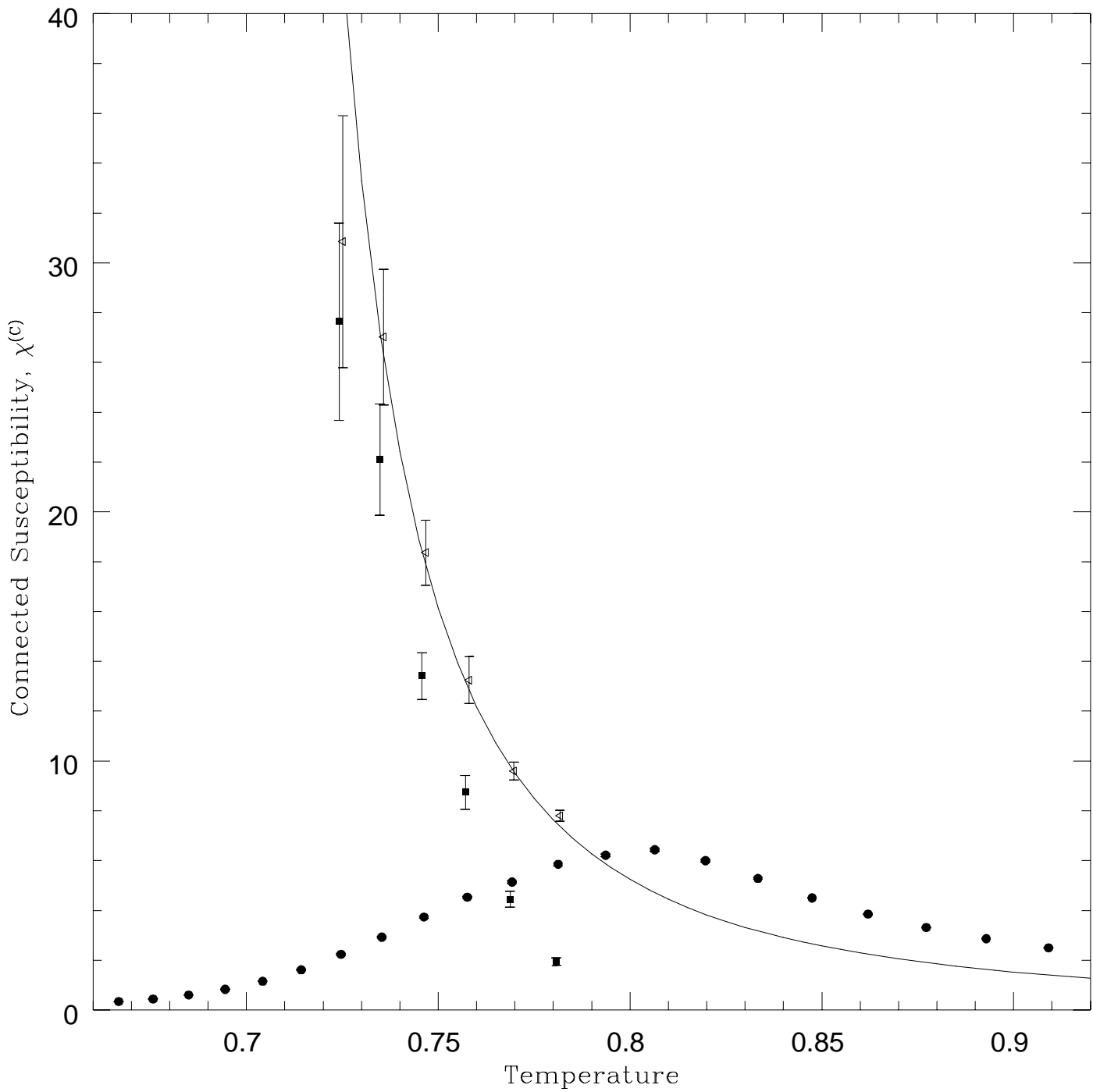
\bye